\documentclass[12pt]{article}
\usepackage{amsmath,latexsym,epsf}
\usepackage{epsfig,amsfonts,cite}
\usepackage{amssymb}
\usepackage{graphicx}

\def\tmod{|\tau|}

\def\Eq#1{Eq.~(\ref{#1})}
\def\Eqs#1#2{Eqs.~(\ref{#1}-\ref{#2})}
\def\EqsAnd#1#2{Eqs.~(\ref{#1}) and (\ref{#2})}

\begin{document}

\pagestyle{myheadings}

\title{High-precision determination of universal amplitude ratios for the $q=3$
  Potts model in $2d$}
\date{\today}
\author{
  {Lev N. Shchur$^{*,**}$, Bertrand Berche$^{**}$ and Paolo Butera$^{***}$}\\
  {\small\it $^*$ Landau Institute for Theoretical Physics, }  \\[-0.2cm]
  {\small\it Russian Academy of Sciences,}  \\[-0.2cm]
  {\small\it Chernogolovka 142432, Russia}  \\[-0.2cm]
  \\
  {\small\it $^{**}$ Laboratoire de Physique des Mat\'eriaux, }  \\[-0.2cm]
  {\small\it Universit\'e Henri Poincar\'e, Nancy 1}     \\[-0.2cm]
  {\small\it BP 239, F-54506 Vand\oe uvre les Nancy Cedex, France} \\[-0.2cm]
  \\
  {\small\it $^{***}$ Istituto Nazionale di Fisica Nucleare, }  \\[-0.2cm]
  {\small\it  Sezione di Milano-Bicocca, }  \\[-0.2cm]
  {\small\it Piazza delle Scienze 3, 20126, Milano,   Italia}\\[-0.2cm]
  \\
  {\small {\tt lev@landau.ac.ru, }}\\[-0.2cm]
  {\small {\tt berche@lpm.u-nancy.fr, }}\\[-0.2cm]
  {\small {\tt paolo.butera@mib.infn.it}}\\[-0.2cm]
  {\protect\makebox[5in]{\quad}}  
  \\
}
\vspace{0.5cm}

\newcommand{\be}{\begin{equation}}
\newcommand{\ee}{\end{equation}}
\newcommand{\bey}{\begin{eqnarray}}
\newcommand{\eey}{\end{eqnarray}}
\newcommand{\<}{\langle}
\renewcommand{\>}{\rangle}

\maketitle
\thispagestyle{empty}   

\vspace{0.2cm}

\begin{abstract}

 Monte Carlo (MC) simulations and series expansion (SE) data for the
energy, specific heat, magnetization and susceptibility of the 3-state
Potts model on the square lattice are analyzed in the vicinity of the
critical point in order to estimate universal combinations of critical
amplitudes.  We estimate these amplitudes
using the correction-to-scaling exponents predicted by conformal field
theory.  We also form effective ratios of the observables close to
the critical point and analyze how they approach the universal
critical-amplitude ratios. In particular, using the duality relation,
we show analytically that for the Potts model with $q \leq 4$ 
states, the effective ratio of the energy critical amplitudes
always approaches unity  linearly with respect to the reduced
temperature. This fact leads to the prediction of relations among the
amplitudes of correction-to-scaling terms of the specific heat in the
low- and high-temperature phases. We present numerical and analytical
support for the form of the first two correction-to-scaling terms. Our
results for the amplitude ratios closely agree with the theoretical
predictions and the earlier numerical estimates of the specific-heat
and the susceptibility amplitude-ratios.

\end{abstract}

%
%

\section{Introduction}
\label{sec-intro}

The universal character of appropriate combinations of critical
amplitudes is an important prediction of scaling theory which  in
some cases still remains incompletely verified and subject to
controversies.
The concept of universality is very fruitful for the classification of
the models and of the real physical systems undergoing phase
transitions.  The set of critical exponents and critical amplitudes
describes the behavior of a system in the vicinity of the critical
point. The universal combinations of critical amplitudes, together
with the critical exponents, fully characterize the universality
class~\cite{PrivmanHohenbergAharony91}. The Potts
model~\cite{Potts52,Wu82}, one of the paradigmatic models exhibiting
continuous phase transitions is a good frame to reconsider the
question of universal combinations of amplitudes. The universality
class of the Potts model at its critical point is determined by the
number of states $q$. The two-dimensional Potts models with three and
four states can be experimentally realized as strongly chemisorbed
atomic adsorbates on metallic surfaces at sub-monolayer
concentrations~\cite{SokolowskiPfnuer94} in accordance with the
group-theoretical classification of the phase transitions of such
systems~\cite{DomanyEtAl78,DomanySchick79,Rottman81}.  Although
critical exponents can be measured quite accurately for adsorbed
sub-monolayers, confirming that these systems actually belong to the
three-state~\cite{NakajimaEtAl97} or to the four-state Potts model
classes~\cite{VogesPfnuer98}, it is unlikely that the low temperature
LEED methods can be pushed~\cite{PfnuerPrivateComm} to determine also
the critical amplitudes. Therefore, the numerical analysis of these
models is the only available tool to check analytic predictions.

We shall restrict our analysis to the critical amplitudes and critical
exponents which describe the behavior of the residual magnetization
$M$, the reduced susceptibility $\chi$, and the reduced specific heat
$C$ of the system at zero external field in the vicinity of the
critical point

\begin{eqnarray}
    M(\tau) &\approx& B (-\tau)^\beta ,\ \tau <0 \label{m-crit}\\
  \chi_\pm(\tau) &\approx& \Gamma_\pm \tmod^{-\gamma}, \label{x-crit}\\
C_\pm(\tau) &\approx& \frac{A_\pm}{\alpha}\tmod^{-\alpha}.
    \label{c-crit}
\end{eqnarray}
Here $\tau$ is the reduced temperature $\tau=(T-T_c)/T$ and the labels
$\pm$ refer to the high-temperature and low-temperature sides of the
critical temperature $T_c$. In addition to the above quantities, for
the Potts models with $q>2$ a transverse susceptibility can be defined
in the low-temperature phase\footnote{In the following we will use the
notations $\Gamma_L$ and $\Gamma_T$ for the longitudinal and
transverse susceptibility amplitudes in the low temperature phase.
$\Gamma_-$ also used in the literature is identified to $\Gamma_L$.}
\begin{equation}
\chi_T(\tau) \approx \Gamma_T (-\tau)^{-\gamma}.
    \label{xt-crit}
\end{equation}

For the 2D Potts model the critical exponents have been calculated
 exactly ~\cite{denNijs79,Pearson80,Nienhuis84,DotsenkoFateev84} in
 terms of the number of states $q$. Introducing the parameter $y$
 related to the number of states $q$ of the model by
\begin{equation}
    \cos\frac{\pi y}{2}=\frac12 \sqrt{q}\,.
    \label{y-from-q}
\end{equation}
we have for the thermal scaling dimension $x_\epsilon=(1-\alpha)/\nu$
\begin{equation}
    x_\epsilon=\frac{1+y}{2-y}
    \label{x-T}
\end{equation}
and for the  magnetic scaling dimension $x_\sigma=\beta/\nu$
\begin{equation}
    x_\sigma=\frac{1-y^2}{4(2-y)}\,.
    \label{x-m}
\end{equation}

The central charge of the corresponding conformal field theory is
also simply expressed~\cite{DotsenkoFateev84} in terms of $y$
\begin{equation}
    c=1-\frac{3y^2}{2-y}\,.
    \label{central-c}
\end{equation}

Analytical estimates of critical amplitude-ratios for the $q$-state
Potts models with $q=1$, $2$, $3$, and $4$ were obtained by Delfino
and Cardy~\cite{DelfinoCardy98}. They used the exact two-dimensional
scattering field theory of Chim and
Zamolodchikov~\cite{ChimZamolodchikov92} and estimated the ratios
using a two-kink approximation for $1<q\le 3$ and the contributions
from both the two-kink approximation and the bound state for
$3<q\le4$.  This approximation leads to the value $c=0.792$ for the
central charge of the 3-state Potts model, for which the exactly known
value is $c=4/5$. Thus, the value of the central charge $c$ is
reproduced with an accuracy of one per cent. Using this approximate
value in (\ref{central-c}), we can calculate the scaling dimensions
from (\ref{x-T})-(\ref{x-m}) obtaining  $x_\sigma=0.1332$
and $x_\epsilon=0.806$, which differ from the exactly known dimensions
($2/15$ and $4/5$ respectively) by less than one per cent.

The universal susceptibility amplitude-ratios $\Gamma_+/\Gamma_L$
were calculated by Delfino and Cardy in~\cite{DelfinoCardy98}. Later
Delfino, et al.~\cite{DelfinoBarkemaCardy00} estimated analytically
also the ratio of the transverse to the longitudinal susceptibility
amplitude $\Gamma_T/\Gamma_L$. They obtained:
\begin{equation}
\Gamma_+/\Gamma_L\approx 13.848, \;\;\; \Gamma_T/\Gamma_L\approx
0.327 \,.
    \label{res-theor-q3}\\
\end{equation}

\noindent In the same paper~\cite{DelfinoBarkemaCardy00}, MC
 simulations were also reported which yield
\begin{equation}
\Gamma_+/\Gamma_L\approx 10, \;\;\;  \Gamma_T/\Gamma_L = 0.333(7).
    \label{res-DBC-q3}
\end{equation}

In our previous paper~\cite{ShchurButeraBerche02}, from the analysis
of MC data and of old SE data for the 3-state Potts model, we
estimated $\Gamma_+/\Gamma_L=14{\pm}1$ in fair agreement with the
theoretical prediction. The key point of our analysis was a fit
including a correction-to-scaling term. Quite recently, these results
were confirmed and substantially improved by Enting and
Guttmann~\cite{EntingGuttmann03} who analyzed new longer series
expansions derived by the finite lattice method. Their remarkably
accurate estimates for the 3-state Potts model are:
\begin{equation}
\Gamma_+/\Gamma_L=13.83(9)-13.90(15), \;\;\; 
\Gamma_T/\Gamma_L=0.325(2)-0.329(2)\,.
    \label{res-EG-q3}
\end{equation}

In the present paper, devoted to the 3-state Potts model, we present
new - more accurate - MC data supplemented by a reanalysis of the
extended series derived by Enting and
Guttmann~\cite{EntingGuttmann03}. We also present numerical and analytical
support for the form of the first two correction-to-scaling terms.

We shall be concerned with the following universal combinations
 of critical amplitudes
\begin{equation}
    \frac{A_+}{A_-}, \;\; \frac{\Gamma_+}{\Gamma_L}, \;\;
    \frac{\Gamma_T}{\Gamma_L}, \;\; R_C^+=\frac{A_+\Gamma_+}{B^2},
    \;\; R_C^-=\frac{A_-\Gamma_-}{B^2},
    \label{univ-rat}
\end{equation}
where the last two are a consequence of the scaling
relation\footnote{We refer the reader to
Ref.~\cite{PrivmanHohenbergAharony91} for a detailed discussion of the
universality of the critical amplitude ratios.}
$\alpha=2-2\beta-\gamma$. To the various critical amplitudes of
interest, $A_\pm$, $\Gamma_\pm$,\dots, we have associated
appropriately defined ``effective amplitudes'', namely
temperature-dependent quantities $A_\pm(\tmod)$,
$\Gamma_\pm(\tmod),\ldots$, which take as limiting values, when
$\tmod \rightarrow 0$, the critical amplitudes
$A_\pm$, $\Gamma_\pm,\dots$.
By analogy, we also considered the ``effective ratios'' of some
amplitudes, e.g., $A_+(\tau)/A_-(-\tau)$ which takes in the critical
limit the value of the first amplitude ratio in the
list~(\ref{univ-rat}). To avoid any risk of confusion, reference to
these temperature-dependent quantities is always made with their
explicit $\tau-$dependence.

A central idea in our approach is to use the duality relation in order
to improve the estimates of the effective amplitude-ratios measured at
{\em dual} temperatures. In a first analysis, we also study directly the
ratios of the effective amplitudes in~(\ref{univ-rat}) at symmetric
reduced temperatures $\pm\tau$ above and below $T_c$. Better than
separately estimating high- and low-temperature effective amplitudes,
these effective ratios enable us to minimize correction-to-scaling
effects in the accessible critical window.  We also use the duality
relation to estimate the leading correction-to-scaling amplitudes in the
behavior of the specific heat and of the susceptibility. For this
purpose, we compute ratios also on the duality line, e.g., the
effective susceptibility amplitude ratio
$\Gamma_+(\tau)/\Gamma_L(\tau^*) =\chi_+(\beta)/\chi_L(\beta^*)$ as
the ratio of $\chi_+(\beta)$, the high-temperature susceptibility at
the inverse temperature $\beta$, and $\chi_L(\beta^*)$, the
low-temperature susceptibility at the dual inverse temperature
$\beta^*$.

As a final result of our analysis, we estimate the susceptibility
critical amplitude-ratios which, for $q=3$, take the values
$\Gamma_T/\Gamma_L=0.3272(2)$ and $\Gamma_+/\Gamma_L=13.83(8)$.
These ratios are consistent with our previous results and with the
predictions of Ref.~\cite{DelfinoCardy98}.

%
%
\section{Model and technical details}
\label{sec-model}

The Hamiltonian of the Potts model~\cite{Potts52} reads as
\begin{equation}
    H = - \sum_{\langle ij \rangle}\delta_{s_i s_j}\; ,
    \label{Ham}
\end{equation}
where $s_i$ is a ``spin'' variable taking integer values between $0$
and $q-1$, and the sum is restricted to the nearest neighbor sites
$\langle ij \rangle$  on a  lattice of $N$ sites with periodic
boundary conditions. The partition function $Z$ is defined as usual
by the sum over all spin configurations
\begin{equation}
    Z =  \sum_{conf} {\rm e}^{-\beta H}.
    \label{Z-part}
\end{equation}
 with $\beta=1/k_BT$, and $k_B$ the Boltzmann constant (fixed to unity).
On the square lattice in zero magnetic field, the model is
self-dual. Denoting by $\beta^*$ the dual of the inverse temperature
$\beta$, the  duality relation~\cite{Potts52}
\begin{equation}
    \left( e^\beta - 1 \right) \left( e^{\beta^*} - 1 \right)=q
    \label{d-t}
\end{equation}
determines  the critical value of the inverse temperature
$\beta_c=\ln (1+\sqrt{q})$. From the duality transformation of the
partition function $Z(\beta)=(q^{-1/2}(e^\beta
-1))^{2N}q^{-1}Z(\beta^*)$, a similar transformation follows for the
internal energy density $E(\beta)=-N^{-1}\frac{d}{d\beta}\ln Z(\beta)$. The
values $E(\beta)$ and $E(\beta^*)$ of the  energy density at dual
temperatures are thus related through (see, e.g.,
Ref.~\cite{CaselleTateoVinci99})
\begin{equation}
    \left(1-e^{-\beta}\right) E(\beta) + \left(1-e^{-\beta^*}\right)
    E(\beta^*)=-2.
    \label{d-et}
\end{equation}

Dual reduced temperatures $\tau$ and $\tau^*$ can be defined by
$\beta=\beta_c(1-\tau)$ and $\beta^*=\beta_c(1+\tau^*)$.
Close to the critical point, $\tau$ and
$\tau^*$ coincide through the first  order, since $\tau^*=\tau+\frac{\ln
(1+\sqrt q)}{\sqrt q}\tau^2+O(\tau^3)$.

%
%
\subsection{Monte Carlo simulations}
\label{sec-MC}

We use the Wolff algorithm~\cite{Wolff89} for studying square
lattices of linear size $L$ with periodic boundary conditions.
Starting from an ordered state, we let the system equilibrate in
$10^5$---$10^6$ steps measured by the number of flipped Wolff
clusters. The averages are computed over $10^6$---$10^7$ steps. The
random numbers are produced by an exclusive-XOR combination of two
shift-register generators with the taps (9689,471) and (4423,1393),
which are known~\cite{Shchur99} to be safe for the Wolff algorithm.

The order parameter of a microstate
${\tt M}({\tt t})$ is evaluated during the simulations as
\begin{equation}
    {\tt M}=\frac{qN_m/N-1}{q-1},
    \label{Order-Potts}
\end{equation}
where $N_m$ is the number of sites $i$ with $s_i=m$ at the time
$\tt t$ of the simulation~\cite{Binder81}, and $m\in [0,1,...,(q-1)]$
is the spin value  of the majority of the spins. $N=L^2$ is the
total number of spins. The thermal average is denoted by $M=\<{\tt M}\>$.

Thus, the reduced longitudinal susceptibility in the low-temperature
phase is measured by the fluctuation of the majority-spin
orientation
\begin{equation}
    \beta^{-1}\;\chi_L=\langle N_m^2\rangle-\langle N_m\rangle^2
    \label{susc-LT}
\end{equation}
and the reduced transverse susceptibility is defined in the
low-temperature phase as the fluctuation of the minority of the
spins
\begin{equation}
    \beta^{-1}\;\chi_T=\frac{1}{q-1}\sum_{\mu\ne m}
    (\langle N_\mu^2\rangle-\langle N_\mu\rangle^2),
    \label{susc-T}
\end{equation}
while in the high-temperature phase $\chi_+$ is given by the fluctuations
in all $q$ states,
\begin{equation}
    \beta^{-1}\;\chi_+=\frac{1}{q}\sum_{\mu=0}^{q-1}
    (\langle N_\mu^2\rangle - \langle N_\mu\rangle^2),
    \label{susc-HT}
\end{equation}
where $N_\mu$ is the number of sites with the spin in the state $\mu$.
Properly allowing for the finite-size effects, this definition of the
susceptibility is, in both phases, completely consistent with the
available series expansion data~\cite{ShchurButeraBerche02}.

The internal energy density of a microstate is calculated as
\begin{equation}
    {\tt E}=-\frac{1}{N} \sum_{\langle ij \rangle}\delta_{s_i s_j}\,
    \label{energy}
\end{equation}
its ensemble average is denoted by $E=\<{\tt E}\>$ and the reduced
specific heat per spin is given by the energy fluctuations,
\begin{equation}
   \beta^{-2}\; C=-\frac{\partial  E }{\partial\beta}
    =\left(\langle {\tt E}^2 \rangle - \langle {\tt E} \rangle^2 \right)N.
    \label{heat}
\end{equation}

We have simulated the model on square lattices with linear sizes
$L=20$, $40$, $60$, $80$, $100$ and $200$.  In each case, we have
measured the physical quantities in a range of reduced temperatures
called the ``critical window'' and defined as follows. Assuming a
proportionality factor of order 1 in the definition of the correlation
length, the relation $L \approx \xi \propto \tmod^{-\nu}$ yields the
value of the reduced temperature at which the correlation length
becomes comparable with the system size $L$ and thus below which the
finite-size effects are not negligible.  This value defines the lower
end of the critical window. For example in the $q=3$ case, for systems
of sizes $L=100$ and $L=200$, we have $\tmod_{min}(L=100) \approx
0.004$ and $\tmod_{min}(L=200) \approx 0.0017$, respectively. The
upper limit of the critical window is fixed for convenience as the
value of the reduced temperature up to which the corrections to
scaling in the Wegner asymptotic expansion~\cite{Wegner72} 
do not exceed a few
percent, say $2-3 \%$, of the leading critical behavior
\Eqs{m-crit}{c-crit}.

%
%
%
\subsection{Series expansions}
\label{sec-SE}

Our MC study of the critical amplitudes will be supplemented by an
analysis of the high-temperature (HT) and low-temperature (LT)
expansions for $q=3$ recently calculated through remarkably high
orders by Briggs, Enting and 
Guttmann~\cite{BriggsEntingGuttmann94,EntingGuttmann03}. In terms of
these series, we can compute the effective critical amplitudes for the
susceptibilities, the specific heat and the magnetization and
extrapolate them by the current resummation techniques, namely simple
Pad\'e approximants (PA) and differential approximants (DA) properly
biased with the exactly known critical temperatures and critical
exponents.

The LT expansions, expressed in terms of the variable $z= e^{-\beta}$,
extend through $z^{46}$ in the case of the energy. The expansion of
the longitudinal susceptibility extends through $z^{71}$.  In the case
of the transverse susceptibility the corresponding order is $z^{53}$.
The magnetization expansion extends through $z^{47}$.

The HT expansions, computed in terms of the variable
$v=(1-z)/(1+(q-1)z)$, extend to $v^{46}$ in the case of the energy,
 and up to $v^{28}$ in the case of the susceptibility.

It is useful to point out that, for convenience, in
Ref.~\cite{EntingGuttmann03} the product of the susceptibility by the
factor $q^2/(q-1)$, rather than the susceptibility itself, has been
tabulated at HT because it has integer expansion coefficients. For the
same reason, at LT the magnetization times $q/(q-1)$ has been
tabulated.  Therefore the appropriate normalizations should be
restored in order that the series yield amplitudes consistent with the
MC results.

As a general remark on our series analysis, we may point out that in
the $q=3$ case, the accuracy of the amplitude estimates is good, due to
the relatively harmless nature of the power-like corrections to
scaling.

%
\section{Critical amplitudes and universal ratios}
\label{sec-a-3}
\subsection{Expected temperature-dependence of the observables}

In the vicinity of the critical point, the reduced specific heat is
generally expected to behave as
\begin{equation}
    C(\tau)=\frac {A_\pm}{\alpha} \tmod^{-\alpha}
    {\cal F}_{corr}(\tmod^\Delta)+{\cal G}_{bt}(\tau),
    \label{C3} \end{equation}
where ${\cal F}_{corr}(\tmod^\Delta)$ is the correction-to-scaling
function and ${\cal G}_{bt}(\tau)$ represents an analytic background
($bt$ here stands for ``background term'') which accounts for
non-singular contributions to the specific heat, i.e. ${\cal
G}_{bt}(\tau)=D_\pm+D'_\pm\tmod+\dots$. The specific-heat and the
leading correction-to-scaling exponents for $q=3$ are given by
$\alpha=1/3$ and $\Delta=-\nu (D-x_{\epsilon_2})=2/3$ where $D=2$ is
the space dimension~\cite{Nienhuis82} and
$x_{\epsilon_2}=(4+2y)/(2-y)$ is the next-to-leading thermal exponent.
In the HT phase and in the LT phase respectively, we can thus
write\footnote{Here and in the following, unless the contrary is
stated, $\tau$ is defined as {\em positive}, but when the physical
quantities are measured in the LT phase, $\tau=-|\tau|$, their
temperature-dependence is explicitly denoted as $Q(-|\tau |)$.}

\begin{eqnarray}
    C_+(\tau)&=&\frac {A_{+}}{\alpha} \tau^{-\alpha}
    \left( 1+ a_{\Delta, +} \tau^{\Delta}
    + b_{+}\tau  + \ldots   \right) + D_+ +D'_+\tau +\dots
    \label{C+}\\
    C_-(-|\tau|)&=&\frac {A_{-}}{\alpha} \tmod^{-\alpha}
    \left( 1+ a_{\Delta, -} \tmod^{\Delta}
    + b_{-}\tmod + \ldots   \right) + D_- + D'_-\tmod+\dots
    \label{C-}
\end{eqnarray}
where $a_{\Delta,\pm}$ is the amplitude of the leading
correction-to-scaling, $b_\pm$ is the next correction term and so
on. In the correction-to-scaling factor the ellipse denotes terms in
$|\tau|^{2\Delta}$, $|\tau|^{3\Delta}$, \dots, and also terms with
an increasing sequence of other exponents~\cite{DotsenkoFateev84}
$\Delta'=-\nu (D-x_{\epsilon_3})=5\Delta$, $\Delta''=-\nu
(D-x_{\epsilon_4})= 14\Delta$, etc. Other quantities should obey
similar expansions including, beside the leading singularity, all
corrections to scaling and background corrections, \bey
    {\rm Obs.}(\pm|\tau|)
        &\simeq&{\rm Ampl.}\times
    \tmod^{\blacktriangleleft}\times
    (1+{\rm corr.\ terms})+
    {\rm\  backgr.\ terms},
        \label{eq-Obs_q3}\\
    {\rm corr.\ terms}&=&a \tmod^{2/3}+b\tmod+\dots,\\
    {\rm backgr.\ terms}&=&D_0+D_1\tmod+\dots
\eey
where ${\blacktriangleleft}$ is a critical exponent which is
known and depends on
the observable considered.

\subsection{Specific-heat critical amplitudes}

Although duality determines exactly the ratio of the specific-heat
amplitudes $A_+/A_-=1$, it is instructive to define an efficient
numerical procedure to compute this ratio. Moreover, the actual
values of the amplitudes $A_+$ and $A_-$, albeit non-universal, are
themselves informative, since they enter into other universal
combinations, e.g. $R_c^{\pm}=A_{\pm}\Gamma_{\pm}/B^2$.

%
%
\subsubsection{Corrections to scaling and background terms}

It is not convenient to extract the critical amplitudes $A_\pm$
directly from the specific heat. The energy density can
be measured more accurately in the MC simulations, so in the
following we shall study
the dominant corrections to scaling and extract the
background term from an analysis of the energy density.

In the HT phase and
in the LT phase respectively, the energy can be conveniently written
as

\begin{eqnarray}
    E_{+}(\tau)&=&E_0 + \frac {A_{+}}{\alpha(1-\alpha)\beta_c} \tau^{1-\alpha}
    \left( 1+ a_{\Delta, +} \tau^{\Delta}
    + b_{+}\tau  + \ldots   \right) + D_+\tau +\dots
    \label{E+}\\
    E_{-}(-|\tau|)&=&E_0 - \frac {A_{-}}{\alpha(1-\alpha)\beta_c}
   \tmod^{1-\alpha}
    \left( 1+ a_{\Delta, -} \tmod^{\Delta}
    + b_{-}\tmod + \ldots   \right) + D_-\tmod+\dots
    \label{E-}
\end{eqnarray}
The minus sign in front of $A_-/\alpha(1-\alpha)\beta_c$ is needed in order to
recover, from the definition $C_-(-\tmod)=\beta_c \partial E_-/\partial\tau=
-\beta_c \partial E_-/\partial\tmod$, the convenient specific heat amplitude
$+A_-/\alpha$.
The
last term represents the analytic background which may be
rewritten as $(A_\pm/\alpha(1-\alpha)\beta_c
)\tmod^{1-\alpha}d_\pm\tmod^{\alpha}+\dots$, in order to be
incorporated in the first sum,
\begin{equation}
    E_{\pm}(\pm\tmod)=E_0 \pm \frac {A_{\pm}}{\alpha(1-\alpha)\beta_c}
    \tmod^{1-\alpha}
    \left( 1+ a_{\Delta,\pm} \tmod^{\Delta}
    + b_{\pm}\tmod  + \ldots + d_\pm\tmod^{\alpha}+\dots  \right).
    \label{E+-}
\end{equation}
%

\begin{figure}[ht]
  \centering
  \begin{minipage}{\textwidth}
  \epsfig{file=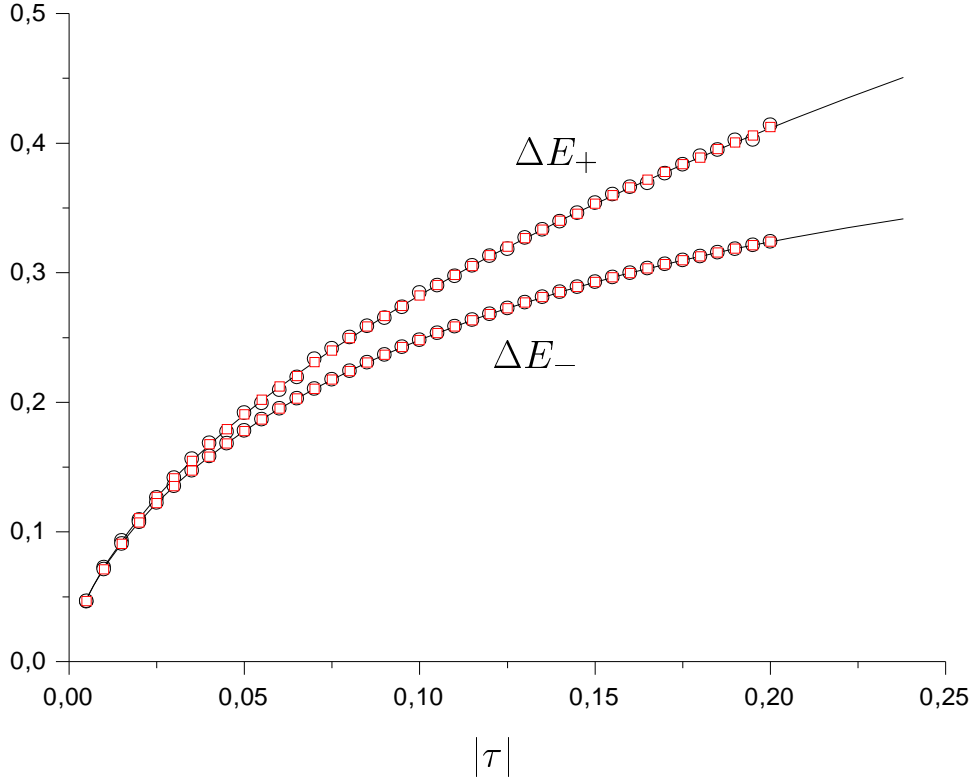,width=0.80\textwidth}
  \end{minipage}
  \vspace{-5cm}
  \caption{\small 
    Energy differences $\Delta E_+$ and
    $\Delta E_-$, calculated from MC data for system sizes $L=100$
    (squares) and $L=200$ (circles)  and from SE data (continuous lines).
    \label{e-sym-3p-full}
    }
\end{figure}

In Fig.~\ref{e-sym-3p-full} we have plotted the differences $\Delta
E_+=E_+(\tau)-E_0$ and $\Delta E_-=E_0-E_-(-|\tau|)$ vs. the reduced
temperature $\tau$. These quantities are computed from both MC data
(symbols) and SE (lines).  By definition of the critical window, the
finite-size corrections of the MC data are negligible in the range of
reduced temperatures under study.  Let us now define the effective
amplitudes $A_+(\tau)$ and $A_-(-|\tau|)$ evaluated at symmetric
reduced temperatures
\begin{eqnarray}
    A_+(\tau)&=&\alpha(1-\alpha)\beta_c
    (E_+(\tau)-E_0)\tau^{\alpha-1},\label{ef-ampl-3a} \\
    A_-(-|\tau|)&=&\alpha(1-\alpha)\beta_c(E_0-E_-(-|\tau|))\tmod^{\alpha-1}.
    \label{ef-ampl-3b}
\end{eqnarray}
The re-summation of the series expansions of these effective
amplitudes can be performed in various ways, all of which show a good
convergence.  We can compute each amplitude by simple PA's after
performing the variable transformations $w = 1-(1-z/z_c)^{\Delta} $ in
the LT case (or $w' = 1-(1-v/v_c)^{\Delta} $ in the HT case) in order
to allow for the leading corrections to scaling.  Completely
consistent results are obtained by computing first-order inhomogeneous
DA's of the amplitudes directly in the natural variables $z$ and $v$.

According to \Eq{E+-},
the arithmetic mean of the effective amplitudes $A_+(\tau)$ and
$A_-(-|\tau|)$
\begin{equation}
    \bar A(\tau)=\frac 12(A_+(\tau) + A_-(-|\tau|))
    \label{half-ampl-3}
\end{equation}
is expected to behave as
\begin{equation}
    \bar A(\tau)= A \left( 1+ \frac {A_+d_++A_-d_-}{2A}\tau^\alpha
     \right) +O(\tau^\Delta)
    \label{e'}
\end{equation}
where $A=\frac 12(A_++A_-)$ and the  correction term comes from the
leading correction to scaling in \Eq{E+-}.

 \begin{figure}[ht]
  \centering
  \begin{minipage}{\textwidth}
\epsfig{file=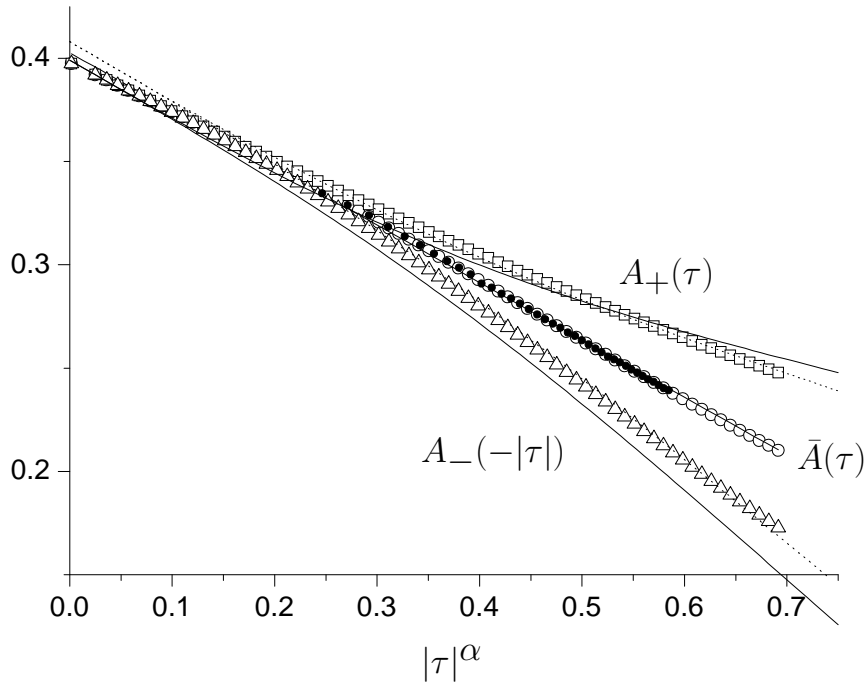,width=0.70\textwidth}
 \end{minipage}
  \vspace{-5cm}
  \caption{\small 
   $A_+(\tau)=\alpha(1-\alpha)\beta_c \Delta E_+\tau^{\alpha-1}$ (open
   squares) and $A_-(-|\tau|)=\alpha(1-\alpha)\beta_c \Delta
   E_-|\tau|^{\alpha-1}$ (open triangles).  Their arithmetic mean
   $\bar A(\tau)$  as computed from SE data (open
   circles), or from MC data(closed circles), together with the fits
   (solid lines: fit of SE data; dotted lines: fit of MC data).
    \label{eee-3s}}
\end{figure}

In order to compare \Eq{e'} with the numerical data, we have plotted
in Figure~\ref{eee-3s} the effective amplitudes $A_+(\tau)$,
$A_-(-|\tau|)$ vs. $|\tau|^{1/3}$.  The SE data for the mean $\bar
A(\tau)$ are well represented asymptotically by the expression
(solid line in Figure~\ref{eee-3s})
\begin{equation}
   \bar A(\tau) \approx 0.399(2)-0.283(1) \tau^{1/3}.
    \label{res-e3}
\end{equation}
for $\tmod^{1/3} > 0.16$ (which corresponds to the left boundary of
the critical window as discussed above). This yields the estimate
$A=0.399(2)$.
We can also conclude from the essentially linear
behavior of $\bar A(\tau)$ with respect to $\tau^\alpha$ that the
higher-order corrections are rather small. Therefore, we can
empirically argue that $A_+a_{\Delta,+}\simeq -A_-a_{\Delta,-}$.
Possibly also a cancellation of some higher order terms might occur
in \Eq{half-ampl-3}. This implies that the most important correction
to
be taken into account comes from the background term.
It is necessary to include the additional terms $\pm 0.15|\tau|^2$
into \Eq{res-e3} (obtained from the  SE data) in order to catch the
behavior at a larger distance from the critical point. This
expression is more accurate also for smaller values of $|\tau|$ (not
accessible through MC due to the finite size effects).

%
%
\subsubsection{The effective amplitude-ratio $A_+(\tau)/A_-(\tau^*)$ on
the dual line}
\label{ef-dual}

The previous empirical observation of a cancellation of several
correction-to-scaling terms in appropriate combinations of effective
amplitudes can be restated more rigorously by duality
arguments. We define effective amplitudes at dual values of the
reduced temperature\footnote{Notice that, with our definitions,
$\tau^*$ is positive and characterizes the LT phase.},
\begin{eqnarray}
    A_+(\tau)&=&\alpha (1-\alpha)\beta_c(E_+(\tau)-E_0)\tau^{\alpha-1}, \label{eff-Aa}\\
    A_-(\tau^*)&=&\alpha (1-\alpha)\beta_c(E_0-E_-(\tau^*))(\tau^*)^{\alpha-1}
    \label{eff-Ab}
\end{eqnarray}
and their ratio

\begin{equation}
\frac{A_+(\tau)}{A_-(\tau^*)}=
\frac{(E_+(\tau)-E_0)\tau^{\alpha-1}}{(E_0-E_-(\tau^*))(\tau^*)^{\alpha-1}}
    \label{rat-e-dual}
\end{equation}
the constant $E_0$ being the value of the energy at the transition
temperature, $E_0=E(\beta_c)=-1-1/\sqrt{q}$.

Using the asymptotic expansions  \EqsAnd{E+}{E-} in the duality
equation \Eq{d-et} and expanding for small $\tau$, we obtain an
infinite set of relations among critical amplitudes,
such as $\quad A_+=A_-$, $\quad D_+=-D_-$,
$\quad a_{\Delta,_+} = a_{\Delta,-}$
$\quad b_-=b_+-2 \alpha_q $.. etc..
These equations hold for any $q \leq 4$.
Therefore

\begin{eqnarray}
    \frac{A_+(\tau)}{A_-(\tau^*)}&=&1+(3-\alpha)\alpha_q \tau
    +O(\tau^{1+\alpha})
    \label{d-e-linear}
\end{eqnarray}
with $\alpha_q=
-E_0\beta_ce^{-\beta_c}=\frac{\ln(1+\sqrt{q})}{\sqrt{q}}=
\approx0.5803$.

Hereafter we shall denote by $A$ the common value of
$A_+$ and $A_-$.

\begin{figure}[ht]
  \centering
  \begin{minipage}{\textwidth}
\epsfig{file=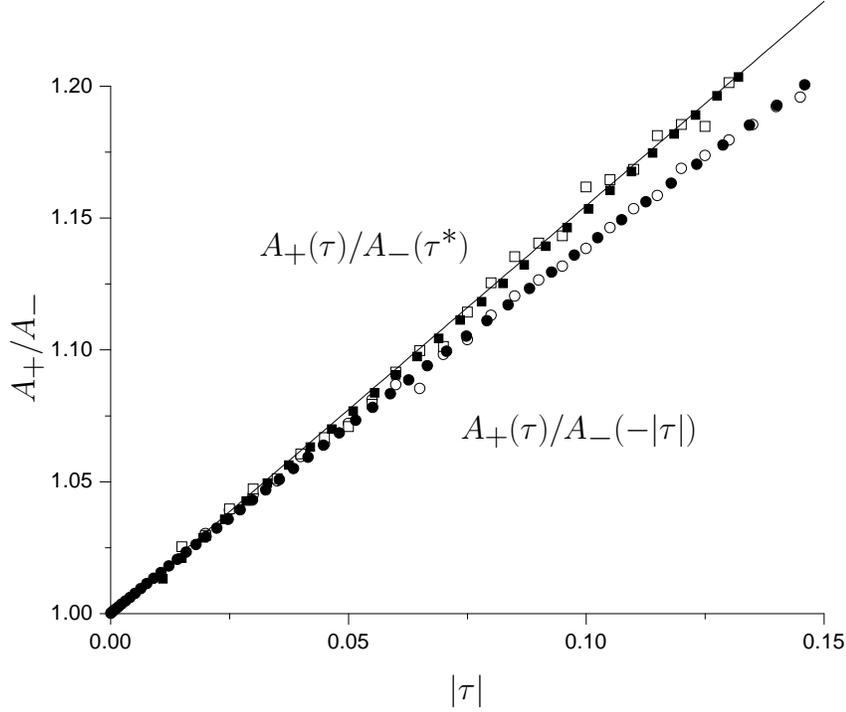,width=0.65\textwidth}
  \end{minipage}
  \vspace{-4.5cm}
  \caption{\small 
    The effective-amplitude ratio
    $A_+(\tau)/A_-(-\tmod)$ at {\em symmetric reduced temperatures}.
    The  SE data are indicated by  closed circles
    and   the MC data by open circles. The  ratio
    $A_+(\tau)/A_-(\tau^*)$ at {\em dual reduced temperatures} is
    denoted by closed
    squares in the case of the SE data and by open squares
        for the MC data.
    The solid line is the
    prediction \Eq{d-e-linear}.
\label{eff-3A-ratio}\label{eee-3-dual}}
\end{figure}

It is interesting to check numerically the validity of \Eq{d-e-linear}
for the asymptotic behavior of the ratio of effective amplitudes
evaluated at {\em dual reduced temperatures}. The ``direct division
method'' of HT and LT series suggested in Ref.\cite{LiuFisher89} is
very effective. It consists in computing the quotient of the $A_+(v)$
and the $A_-(z)$ series after taking $v=z$ (remember that
$v_c=z_c$). This amounts precisely to compute the ratio of the
effective amplitudes at dual temperatures.  The quotient series thus
obtained is resummed by simple PA's or DA's and can be extrapolated
to the critical point obtaining the very accurate estimate
$A_+/A_-=1.000000(3)$.  The results for $A_+(\tau)/A_-(\tau^*)$ shown
in figure~\ref{eff-3A-ratio} are compared with the same ratio computed
as function of the {\em symmetric reduced temperature} $\tmod$,
$A_+(\tau)/A_-(-|\tau|)$.

The fit of MC data at dual temperatures yields a value for the slope
$\approx 1.5$ which is not far from the expected value
$8/3 \alpha_q \approx 1.547$ of \Eq{d-e-linear}.
The estimate of the same quantity in the case
of symmetric reduced temperatures $\tmod$
yields $1.38(1)$.

These plots show the {\em cancellation of the leading non-analytic
correction to scaling in the ratio $A_+/A_-$ of effective
amplitudes}.  We can now take advantage of this remark to improve
the fit of the energy data, presented in the previous subsection.
First, let us construct the mean of the effective amplitudes
evaluated at dual reduced temperatures $\bar A_{\rm dual}(\tau)$.
For small $\tau$ the $SE$ curve has the behavior (in perfect
agreement with \Eq{res-e3})
\begin{equation}
    \bar A_{\rm dual}(\tau)=0.399(1)-0.283(2)\tau^{1/3}
    \label{e'-se-dual}
\end{equation}
As noticed above, the coefficient of the higher-order term
$\tau^{2/3}$ is at least one order of magnitude smaller. In the
linear approximation of the duality relation~(\ref{d-e-linear}),
$A_+(\tmod)/A_-(\tau^*)\approx 1+(3-\alpha)\alpha_q \tau$, one has
\begin{equation}
    A_\pm (\pm \tmod)= \bar A_{\rm dual}(\tau) (1 \pm (3-\alpha)\alpha_q/2 \tau).
    \label{e-pm-solution}
\end{equation}
\noindent Combining  (\ref{e'-se-dual}), (\ref{e-pm-solution}) and
using $(3-\alpha)\alpha_q/2\approx 0.774$, one gets
\be A_\pm (\pm\tmod)
     \approx (0.399(1)-0.283(2) \; \tau^{1/3}) (1 \pm 0.774 \; \tau).
    \label{e+e-}
\ee
These expressions are represented in Fig.~\ref{eee-3s} by the solid
lines.  In order to extend the SE data representation of
\Eq{e'-se-dual} to larger values of $\tmod$, we can add the next
background correction term, estimated to be $\mp 0.15\tau^2$.

\subsubsection{Energy and specific heat temperature dependence}

From Eqs.~(\ref{eff-Aa}), (\ref{eff-Ab}), (\ref{e'-se-dual}) and
(\ref{e-pm-solution}), we get
 a numerical expansion
 of the energy of the 3-state Potts model:
\begin{eqnarray}
    E_+(\tau)-E_0
    &=&\frac{1}{\alpha(1-\alpha)\beta_c}\tau^{1-\alpha}\bar A(\tau)
    (1+(3-\alpha)\alpha_q/2 \tau)
    \nonumber \\
    &=&1.787(5) \tau^{2/3} \left (1+0.774(1)  \tau - 0.412(4)\tau^{4/3}
    \right) - 1.269(9) \tau ,
      \label{fit3e+}\\
    E_0-E_-(-\tmod)
    &=&\frac{1}{\alpha(1-\alpha)\beta_c}|\tau|^{1-\alpha}\bar A(|\tau|)
    (1-(3-\alpha)\alpha_q/2 |\tau|)
    \nonumber \\
    &=&1.787(5) |\tau|^{2/3} \left (1-0.774(1)|\tau| +0.412(4)|\tau|^{4/3}\right)
    - 1.269(9) |\tau|.\qquad
    \label{fit3e} \end{eqnarray}
Note that the regular linear term appears through the combination
of the $\tmod^{1/3}$ correction in Eq.~(\ref{e'-se-dual}) with the
leading singularity in $\tmod^{1-\alpha}$.

As a conclusive test, we have plotted \EqsAnd{fit3e+}{fit3e}
in figure~\ref{e-sym-3p-full}, but they cannot be
distinguished from the solid lines representing SE data.

\begin{figure}[t]
  \centering
\epsfig{file=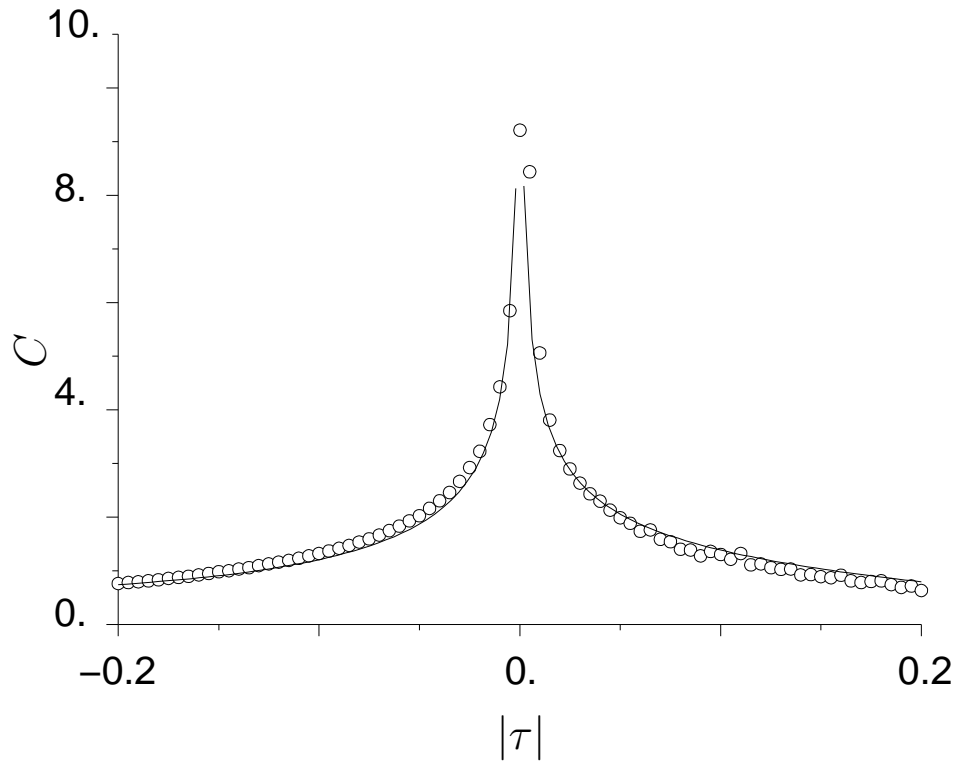,width=0.9\textwidth}\hfill\qquad
\qquad\qquad
  \vspace{-9cm}
  \caption{\small 3-state Potts model. MC data (open circles) and SE data
    (small solid boxes) for specific heat. The continuous line
    represents the specific heat
    calculated from
    \EqsAnd{fit3e+}{fit3e}.}
    \label{sh3}
\end{figure}

We use expressions (\ref{fit3e}-\ref{fit3e+}) to calculate the
specific heat, and plot the results in the figure~\ref{sh3} together
with the MC and SE data.

%
%
\subsection{Susceptibility and magnetization amplitudes}
\label{sec-sus-3}

%
%
\subsubsection{The ratio $\Gamma_T/\Gamma_L$}

The transverse and longitudinal reduced susceptibilities are
expected to have the following asymptotic form in the LT
phase~\cite{Wegner72}
\begin{eqnarray}
    \chi_T(-|\tau|)&=
    &\Gamma_T\tmod^{-\gamma} {\cal F}_T(\tmod^\Delta)+{\cal
    G}_T(\tmod),
    \label{chi-T}\\
    \chi_L(-|\tau|)
    &=&\Gamma_L\tmod^{-\gamma} {\cal F}_L(\tmod^\Delta)+{\cal
    G}_L(\tmod),
    \label{chi-L}
\end{eqnarray}
where $\gamma=13/9$~\cite{Nienhuis82,DotsenkoFateev84} and the
correction-to-scaling exponent $\Delta=2/3$ is the same as above.
For the purpose of the fit in the low temperature range accessible
by our MC simulation, we shall use the following expansion of the
reduced susceptibility
\begin{equation}
    \chi(-|\tau|)=\Gamma \tmod^{-\gamma}
    (1+a_{\Delta}\tmod^\Delta  +a_{2\Delta}\tmod^{2\Delta} +\ldots +
    b \tmod + \ldots
    +d\tmod^\gamma  \ldots \, ),
    \label{chi-tl}
\end{equation}
where for simplicity, we choose a notation for the amplitudes of the
 corrections to scaling very similar to that adopted in the previous
 section, since there is no risk of confusion, and we incorporate the
 leading background term whose amplitude is denoted by $d$ inside the
 main parenthesis.  The ratio of the effective amplitudes
 $\Gamma_T(-|\tau|)=|\tau|^{\gamma}\chi_{T}(-|\tau|)$ and
 $\Gamma_L(-|\tau|)=|\tau|^{\gamma}\chi_{L}(-|\tau|)$ thus behaves as
\begin{eqnarray}
    \frac{\Gamma_T(-|\tau|)}{\Gamma_L(-|\tau|)}
    &=& \frac{\Gamma_T}{\Gamma_L}
    (1+(a_{\Delta, T}-a_{\Delta, L})\tmod^\Delta +
    (b_{T}-b_{L})\tmod +\nonumber\\
    &\phantom{=}&\ \
    (a_{2\Delta, T}-a_{2\Delta, L}+a_{\Delta, L}^2-
    a_{\Delta, T}a_{\Delta, L})
    \tmod^{2\Delta}
    +(d_{T}-d_L)\tmod^\gamma
    +
    O(\tmod^{\Delta +1})).\qquad
    \label{TL-eff}
\end{eqnarray}
 \begin{figure}[ht]
  \centering
  \begin{minipage}{\textwidth}
  \qquad\epsfig{file=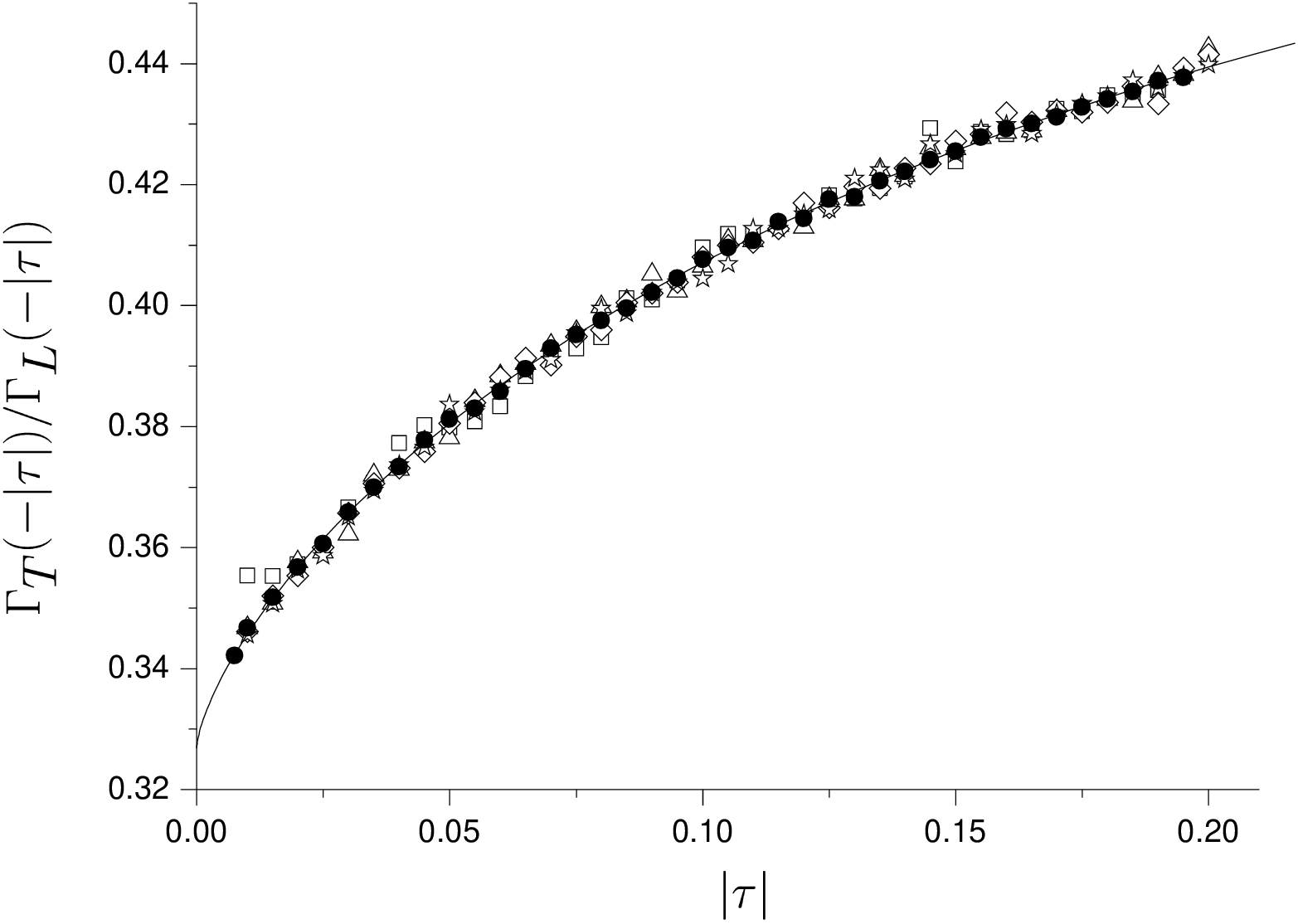,width=0.65\textwidth}
  \end{minipage}
  \vspace{-3.5cm}
  \caption{\small 
  Effective ratio of amplitudes
    $\Gamma_T(-|\tau|)/\Gamma_L(-|\tau|)$ for
    lattices of linear sizes $L=20$ (boxes), $L=40$ (up
    triangles), $L=60$ (down triangles), $L=80$ (diamonds), $L=100$
    (stars) computed with  $N_{MC}=10^5$ Monte Carlo steps.
    We have indicated
    by closed circles the results of a computation on a
    $L=100$ lattice   with $N_{MC}=10^6$  Monte Carlo steps.
    The dashed line (hardly visible) represents a fit
    of the data for $L=100$ to  \Eq{TL-eff}.
    The values of
    the coefficients are listed in the last entry of table~\ref{tab-tl-3}.
    The solid line represents the [22,22] Pad\'e approximant to the ratio
    of the LT series in a conveniently
    transformed  variable.
}
\label{g-eff-3-fs}
\end{figure}
\begin{table}[ht]
\caption{\small Results of the fit to the MC data for the
ratio $\Gamma_T/\Gamma_L$ of the
transverse and the longitudinal susceptibility
in the critical region window.
The absence of an entry in the table indicates
 that we have not included the corresponding
parameter  in the fit procedure.}
\center\scriptsize
\begin{tabular}{llllll} \hline\noalign{\vskip-1pt}\hline\noalign{\vskip2pt}
Fit \# & ratio & \multispan{4}{\hfil corrections coefficients\hfil}\\
&\multispan{1}{\vrule height0.5pt width16mm} &
    \multispan{4}{\vrule height0.5pt width88mm} \\
\noalign{\vskip2pt}
&$\Gamma_T/\Gamma_L$ & $\quad \propto\tmod^\Delta$
    & $\quad\propto\tmod$
    & $\quad\propto\tmod^{2\Delta}$
    & $\quad\propto\tmod^{\gamma}$ 
    \\ \noalign{\vskip2pt}\hline
 1&$0.328(2)$ & $1.04(26)$   & $\phantom{-}1.11(87)$
    & $-2.03(79)$& $\phantom{-}-$
    \\
 2&$0.328(2)$ & $1.13(23)$   & $\phantom{-}0.56(6)$
    & $\phantom{-}-$
    & $-1.63(65)$ \\
 3&$0.327(2)$ & $1.24(12)$   & $\phantom{-}-$
    & $\phantom{-}1.9\pm2.6$& $-3.1\pm2.7$ \\
 4&$0.324(1)$ & $1.71(5)$    & $-1.10(7)$
    & $\phantom{-}-$
    & $\phantom{-}-$
    \\
 5&$0.326(1)$ & $1.37(2)$    & $\phantom{-}-$
    & $-1.02(6)$ & $\phantom{-}-$
    \\
 6&$0.326(1)$ & $1.33(2)$    & $\phantom{-}-$
    & $\phantom{-}-$
    & $-1.08(6)$ \\
\hline\noalign{\vskip-1pt}\hline
\end{tabular}
\label{tab-tl-3}
\end{table}

In Figure~\ref{g-eff-3-fs} we have represented the MC data for this
effective ratio, plotted against the reduced temperature $\tmod$.  The
data in Fig.~\ref{g-eff-3-fs} do not show any finite-size dependence
even for the smallest lattice size, although their spread becomes
obviously smaller for larger system sizes and longer MC runs. We
performed a fit of the MC data collected for $L=100$ and with the
largest statistics (represented by closed circles in
figure~\ref{g-eff-3-fs}) to expression~(\ref{TL-eff}). The values of
the parameters are reported in table~\ref{tab-tl-3} for different trial 
fits.  It is worth
noticing that, while the values of the parameters are sensitive to
variations of the limits of the critical-region window, the amplitude
ratio changes only within the error bars. Therefore the good quality
of the MC data and the large size of the critical-region window enable
us to estimate rather accurately the amplitude and the first
correction term in the asymptotic expansion~(\ref{TL-eff}). In
table~\ref{tab-tl-3}, the leading correction comes from the regular
term (last column) and the coefficient is clearly not negligible,
therefore the fits \# 2, 3 and 6 have our preference. Since it is hard
to decide which one among the three is the most reliable, we consider
that an average result $\Gamma_T/\Gamma_L=0.327(1)$ is a safer value.
Figure~\ref{g-eff-3-fs}, if replotted as a function of $\tmod^{2/3}$
shows a remarkably linear behavior for $\tmod^{2/3}<0.2$, while the
result of the linear fit: $\Gamma_T/\Gamma_L=0.329(1)$ supports our
choice of the value $2/3$ for the leading confluent correction
exponent.

Although possibly less accurate, the MC data for the low temperature
susceptibilities can be fitted in the same window, leading to
\begin{eqnarray}
&\chi_T&=0.00409(2)\tmod^{-13/9}(1+0.32(7)\tmod^{2/3}-1.88(11)\tmod),
\label{chi_T_q3_MC}\\
&\chi_L&=0.01270(4)\tmod^{-13/9}(1-1.60(5)\tmod^{2/3}+0.44(8)\tmod).
\label{chi_L_q3_MC}
\end{eqnarray}
The ratio of the amplitudes, $\Gamma_T/\Gamma_L\approx 0.322$, is
fairly consistent with the value quoted above, but since no background
term is included we consider that this value is not highly
reliable. The differences between coefficients $a_\Delta$'s and $b$'s
appearing in \Eq{chi_T_q3_MC} and (\ref{chi_L_q3_MC}) are also
compatible with those quoted in table~\ref{tab-tl-3}.

The quotient of the LT series for $\Gamma_T$ and $\Gamma_L$ can also
be studied either most simply by PA's in the transformed variable $w$
as mentioned above or by DA's in the variable $z$. Both kinds of
approximants are smoothly extrapolated to the value
$\Gamma_T/\Gamma_L=0.3272(2)$ for the critical amplitude ratio. The
critical amplitudes can also be separately computed by DA's obtaining
the estimates $\Gamma_T=0.004166(5)$ and $\Gamma_L=0.01273(1)$.  Our
results, summarized and compared with those of previous studies in
Table \ref{tab-3-res} and \ref{tab-3-resbis}, agree completely with
the prediction $\Gamma_T/\Gamma_L=0.327$ by Delfino, et
al.~\cite{DelfinoBarkemaCardy00} whose uncertainty is presumably of
the order of a half percent.

%
%
\subsubsection{The critical amplitude of the magnetization.}
The magnetization  is expected to
behave in the critical region as
\begin{equation}
    M(-|\tau|)=B \tmod^{\beta}
    \left(1+a_\Delta\tmod^\Delta + \ldots +
    b\tmod  + \ldots \right) +
    D\tmod  + \ldots,
    \label{M3-exp}
\end{equation}
where  $\beta=1/9$ is  the magnetization exponent and as usually
we have indicated only the leading corrections to scaling and the analytic
background.

\begin{figure}[ht]
  \centering
  \begin{minipage}{\textwidth}
  \epsfig{file=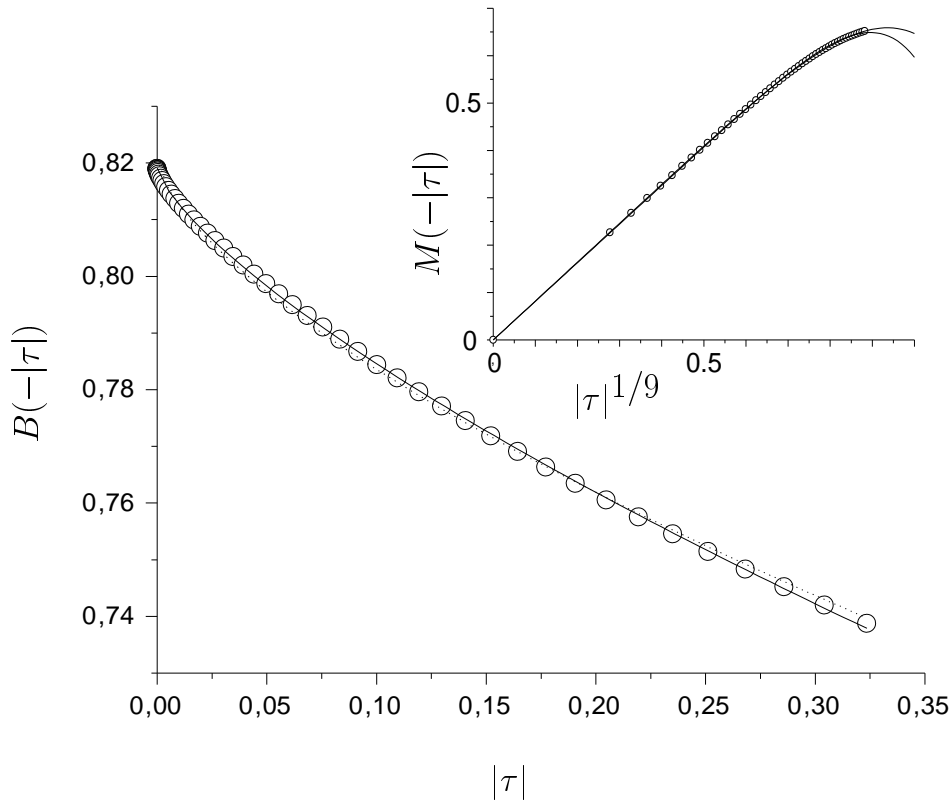,width=0.70\textwidth}
  \end{minipage}
  \vspace{-3.5cm}
\caption{\small 
The effective amplitude of the magnetization
    $M$ as computed from SE (open circles) together with the fits
    (lines, see in text).
    Insert: The magnetization $M$ as function of
    $\tmod^{1/9}$ (open circles) and a fit to the curves $B|\tau|^\beta
    (1+a_M|\tau|^{\Delta_{\rm eff}})$
    with the exponent $\Delta_{\rm eff}$
    fixed to the values $\Delta_{\rm eff}=\Delta=2/3$ (upper line) and
    $\Delta_{\rm eff}=1-\beta=8/9$ (lower line).
}
    \label{b-mag-3}\label{m3}
\end{figure}

The analysis of the LT expansion for the effective amplitude
$B(-|\tau|)=|\tau|^{-\beta}M(-|\tau|)$ of the magnetization
 can be efficiently performed either by simple PA's in the
transformed variable $w$ or by DA's. We obtain the estimate
$B=0.819(1)$ for the critical magnetization amplitude.  DA's also
indicate that the exponent of the leading correction to scaling is
0.73(3),
\begin{equation}
    B(-|\tau|)=0.819(1)-0.226(2)\tmod^{0.73(3)}.
    \label{fit-B-3}
\end{equation}
Comparing (\ref{M3-exp}) to (\ref{fit-B-3}) one can identify
$\tmod^{0.73(3)}$ with $\tmod^\Delta$. As an illustration, a
fit (with the exponent fixed to the value $\Delta=2/3$) of the SE data
is shown in the figure~\ref{m3} (dotted line) and compared with the
fit~(\ref{fit-B-3}) (solid line).

A highly compatible fit is obtained for the MC data which yields\be
M=0.818(1)\tmod^{1/9}(1-0.65(3)\tmod^{2/3}-0.400(7)\tmod+0.2865\tmod^{4/3}).\ee

\subsection{$R_C^\pm$ ratios}

We can finally estimate the value of the universal amplitude ratio
$R_C^+=A_+\Gamma_+/B^2$.  If we use our estimates $A_+=0.399(1)$,
$B=0.819(1)$ and the value $\Gamma_+=0.176(1)$ obtained from the
presently available series\cite{EntingGuttmann03} (much longer than
those used in our previous study\cite{ShchurButeraBerche02}), we
conclude
$$R_C^+=0.1049(8).$$ This value compares very well with the estimate
$R_C^+=0.1041$ by Delfino and Cardy~\cite{DelfinoBarkemaCardy00}.

The corresponding Monte Carlo estimates are
$A_+=0.396(9)$,
$B=0.818(1)$ and  $\Gamma_+=0.1783(7)$ leading to a value for the
ratio
$R_C^+=0.1054(29)$
consistent with SE estimation albeit less
accurate.

An alternative approach, leading to a very similar numerical estimate
$R_C^+=0.1043(8)$, consists in expressing $R_C$ as the combination of
the effective amplitudes $A(v)$, $\Gamma(v)$ and $B(z)$, after taking
$v=z$, namely on the dual line in the $z-v$ plane. This procedure,
however, wastes a large part of the available series coefficients.

Using the value $\Gamma_L=0.01273(1)$ obtained from SE one can
estimate $R_C^-=0.00756(4)$ while using the MC value
$\Gamma_L=0.01270(4)$ one obtains $R_C^-=0.00751(18)$.  These values
agree well with Delfino and Cardy~\cite{DelfinoBarkemaCardy00}
analytic prediction $R_C^-=0.00752$.

\subsection{Summary of the results for the 3-state Potts model}
\label{3-discussion}

By MC simulations and by series extrapolations, we have computed all
universal amplitude-ratios studied in Refs.\cite{DelfinoCardy98} and
\cite{DelfinoBarkemaCardy00} for the $q=3$ Potts model, with the
exception of those involving the correlation length.  We have shown
that the cancellation of the leading non-analytic corrections to
scaling in the ratio of the effective amplitudes of the energy
evaluated at dual temperatures leads to a very accurate estimate of
this amplitude ratio. Using this result, we have given an asymptotic
numerical representation of the energy in a vicinity of the critical
temperature, and observed that the main correction to scaling is due
to the background term.

In table~\ref{tab-3-res} we have collected the results from our fits
of the main observables to the generic form given in \Eq{eq-Obs_q3}.
These fits are performed either using MC data or SE data, and their
results are reported in the last column. From the estimates obtained,
we can form the universal combinations reported in
table~\ref{tab-3-resbis}. We have in general found more accurate
results by fitting {\em effective amplitude ratios} rather than the
amplitudes themselves. A comparison of our numerical results with the
analytical predictions and with numerical estimates from other sources
generally shows a very good agreement.

As a conclusion, we would like to mention that our primary interest
is also to study the universal combinations of amplitudes in the case of
the $4-$state Potts model for which the available results are less convincing
due to the presence of logarithmic corrections to scaling. We believe that
the analysis proposed here in the case of the $3-$state model (essentially 
based on an analysis of quantities at dual temperatures and of effective
temperature-dependent combinations) can be successfully transposed to the 
$4-$state case (see Refs.~\cite{CSP,EPL} for preliminary results).

\begin{table}[t]
\caption{\small Critical amplitudes and correction-to-scaling amplitudes for
    the $3-$state Potts model,
    ${\rm Obs.}(\pm\tmod)\simeq{\rm Ampl.}\times
    \tmod^{\blacktriangleleft}\times
    (1+{\rm corr.\ terms})
    +{\rm backg.\ terms}$.
    The results of the SE data analysis
    of Ref.\cite{EntingGuttmann03} are tabulated together with our results
    obtained  by combining the Monte Carlo and the SE data analysis. Results
	in bold face have our preference.}

\center\scriptsize
\begin{tabular}{lllllllc}
\hline\noalign{\vskip-1pt}\hline\noalign{\vskip2pt}
observable & amplitude & \multispan{3}{\hfil corrections\hfil}
&\multispan{2}{\hfil background\hfil} & source \\
\multispan{1}{\vrule height0pt width0mm} &
\multispan{1}{\vrule height0pt width0mm} &
\multispan{3}{\vrule height0.5pt width48mm} &
\multispan{2}{\vrule height0.5pt width32mm} &
\multispan{1}{\vrule height0pt width0mm}
\\
\noalign{\vskip2pt}
           &           & $\propto\tmod^\Delta$ & $\propto\tmod$ & $\propto
    \tmod^{2\Delta}$
       & $\propto\tmod^0$ &  $\propto\tmod$
       &        \\
\noalign{\vskip2pt} \hline
$E_+(\tau)$ & 
     $\bf \phantom-1.787(5)$
    & $\phantom-\simeq 0$ & $\bf \phantom + 0.774(1)$
    & $\bf -0.412(4)$
    & $\phantom{--}E_0$ & $\bf -1.269(9)$ & this paper$^{c}$ \\
$E_-(-|\tau|)$ & 
     $\bf -1.787(5)$
    & $\phantom-\simeq 0$ & $\bf -0.774(1)$
    & $\bf \phantom + 0.412(4)$
    & $\phantom{--}E_0$ & $\bf \phantom - 1.269(9)$ & this paper$^{c}$ \\
$C_+(\tau)$ & 
$\bf A_+=0.399(1)$ & $\phantom-\simeq 0$ & $\bf \phantom + 1.548(3)$
    & $\bf -1.236(12)$
    & $\bf -1.269(9)$ & $\phantom--$ & this paper$^{b}$ \\
    &
$A_+=0.396(9)$&$\phantom{-}-$&$\phantom{-}-$&$\phantom{-}-$&$\phantom{-}-$&$\phantom{-}-$&this paper$^{a}$ \\
$C_-(-|\tau|)$ &
$\bf A_-=0.399(1)$& $\phantom-\simeq 0$ & $\bf -1.548(3)$ &
    $\bf \phantom + 1.236(12)$
    & $\bf -1.269(9)$ & $\phantom--$ & this paper$^{b}$ \\
        &
$A_-=0.396(9)$&$\phantom{-}-$&$\phantom{-}-$&$\phantom{-}-$&$\phantom{-}-$&$\phantom{-}-$&this paper$^{a}$ \\
$\chi_+(\tau)$ & $\Gamma_+=0.1783(7)$ & $\phantom{-}0.24(2)$
    & $\phantom{-}-$ & $\phantom{-}-$ &
    $\phantom{-}0.005(6)$ & $\phantom{-}-$ &
    \cite{ShchurButeraBerche02} \\
               & $\Gamma_+=0.1751(6)$ & $\phantom{-}-$
    & $\phantom{-}-$ & $\phantom{-}-$
    & $\phantom{-}-$ & $\phantom{-}-$ &
    \cite{EntingGuttmann03} \\
               & $\bf \Gamma_+=0.176 (1)$ & $\phantom{-}-$       &
    $\phantom{-}-$ & $\phantom{-}-$        &
    $\phantom{-}-$ & $\phantom{-}-$ &
    this paper$^{c}$ \\
$\chi_L(-|\tau|)$ & $\Gamma_L=0.012774(3)$ & $-1.517(8)$ & $\phantom{-}-$
    & $\phantom{-}-$ & $\phantom{-}0.0070(2)$ & $\phantom{-}-$ &
    \cite{ShchurButeraBerche02} \\
               & $\Gamma_L=0.01266(4)$ & $\phantom{-}-$
    & $\phantom{-}-$ & $\phantom{-}-$        &
    $\phantom{-}-$ & $\phantom{-}-$ &
    \cite{EntingGuttmann03} \\
               & $\bf \Gamma_L=0.01270 (4)$ & $\bf -1.60(5)$
    & $\bf \phantom{-}0.44(8)$ & $\phantom{-}-$        &
    $\phantom{-}-$ & $\phantom{-}-$ &
    this paper$^{a}$ \\
               & $\Gamma_L=0.01273 (1)$ & $\phantom{-}-$
    & $\phantom{-}-$ & $\phantom{-}-$        &
    $\phantom{-}-$ & $\phantom{-}-$ &
    this paper$^{c}$ \\
$\chi_T(-|\tau|)$ & $\Gamma_T=0.004168(9)$ & $\phantom{-}-$
    & $\phantom{-}-$ & $\phantom{-}-$        &
    $\phantom{-}-$ & $\phantom{-}-$ &
    \cite{EntingGuttmann03} \\
               & $\bf \Gamma_T=0.00409(2)$ & $\bf \phantom{-}0.32(7)$ &
    $\bf -1.88(11)$
    & $\phantom{-}-$ & $\phantom{-}-$ & $\phantom{-}-$
    &
    this paper$^{a}$ \\
               & $\Gamma_T=0.004166 (5)$ & $\phantom{-}-$
    & $\phantom{-}-$ & $\phantom{-}-$        &
    $\phantom{-}-$ & $\phantom{-}-$ &
    this paper$^{c}$ \\
$M(-|\tau|)$ & $B=0.819(1)$ & $\phantom{-}-$ & $\phantom{-}-$
    & $\phantom{-}-$ &
    $\phantom{-}-$ & $\phantom{-}-$ &  this paper$^{c}$ \\
$M(-|\tau|)$ & $\bf B=0.818(1)$ & $\bf -0.65(3)$ & $\bf -0.400(7)$
    & $\bf \phantom{-}0.2865$ &
    $\phantom{-}-$ & $\phantom{-}-$ &  this paper$^{a}$ \\
\hline\noalign{\vskip-1pt}\hline\noalign{\vskip2pt}
\multicolumn{8}{l}{$^a$  fit of MC data}\\
\multicolumn{8}{l}{$^b$  derivative of Eqs.~(\ref{fit3e+})-(\ref{fit3e})}\\
\multicolumn{8}{l}{$^c$  approximants of SE data}\\
\end{tabular}
\label{tab-3-res}
\end{table}

\begin{table}
\caption{\small Universal combinations of the critical amplitudes in 
	the $3-$state
    Potts model. The first line shows the analytical predictions
    of Refs.\cite{DelfinoCardy98} and \cite{DelfinoBarkemaCardy00}.
        The remaining
    lines are obtained  by combining the Monte Carlo and series
    expansion (SE) data analysis (in particular, the second 
    and third lines show the results of the SE data analysis
    of Ref.\cite{EntingGuttmann03}).}
\center\scriptsize
\begin{tabular}{lllllc}
\hline\noalign{\vskip-1pt}\hline\noalign{\vskip2pt} $A_+/A_-$ &
$\Gamma_+/\Gamma_L$ & $\Gamma_T/\Gamma_L$ & $R_C^+$ & $R_C^-$ &
source
\\ \noalign{\vskip2pt}\hline
$1$ & $13.848$ &  $0.327$  & $0.1041$  & 0.00752 &
\cite{DelfinoCardy98}
or \cite{DelfinoBarkemaCardy00} \\
& $13.83(9)$ & $0.325(2)$  & $-$ & - & \cite{EntingGuttmann03} \\
& $13.90(15)$ & $0.329(2)$  & $-$ & - & \cite{EntingGuttmann03} \\
& $14(1)$ & $-$  & $-$ & - & \cite{ShchurButeraBerche02} \\
$1.000(1)$ & $13.86(12)$ &  $ 0.322(3)$
    & $0.1049(29)$ & 0.00748(18)
    & {this paper$^{a}$}\\
$\phantom--$ & $13.83(9)$ &  $0.3272(7)$
    & 0.1044(8) & 0.00753(4)
    & {this paper$^{b}$}\\
$1.000000(3)$ & $\phantom--$ &  $0.327(1)$
    & $0.1038(8)$ & -
    & {this paper$^{c}$}\\
\hline\noalign{\vskip-1pt}\hline\noalign{\vskip2pt}
\multicolumn{5}{l}{$^a$  from amplitudes extracted from fits
of MC data quoted in table~\ref{tab-3-res}}
\\
\multicolumn{5}{l}{$^b$  from amplitudes extracted from fits
of SE data quoted in table~\ref{tab-3-res}}
\\
\multicolumn{5}{l}{$^c$ from effective amplitude ratio, SE data,
PA's or DA's}
\\
\end{tabular}
\label{tab-3-resbis}
\end{table}

\section{Acknowledgements}

LNS is grateful to the Statistical Physics group of the University
Henri Poincar\'e Nancy~1 for the kind hospitality. Both LNS and PB
thank the Theoretical group of the University Milano--Bicocca for
hospitality and support. Financial support from the twin research
program between the Landau Institute and the Ecole Normale
Sup\'erieure de Paris and Russian Foundation for Basic Research are
also acknowledged.

%
%


\end{document}